\documentclass[showkeys,preprintnumbers,amsmath,amssymb,prl,twocolumn,nofootinbib]{revtex4}
\usepackage{epsfig,verbatim}

\begin{document}

\title{Physics Engineering in the Study of the Pioneer Anomaly}

\author{Slava G. Turyshev}
\affiliation{Jet Propulsion Laboratory, California Institute of Technology}
\email{turyshev@jpl.nasa.gov}

\author{Viktor T. Toth}
\address{Ottawa, ON  K1N 9H5, Canada}
\homepage{http://www.vttoth.com/}

\begin{abstract}

The Pioneer 10/11 spacecraft yielded the most precise navigation in deep space to date.  However, their radio-metric tracking data received from the distances between 20--70 astronomical units from the Sun has consistently indicated the presence of a small, anomalous, Doppler frequency drift.  The drift is a blue frequency shift that can be interpreted as a sunward acceleration of $a_P = (8.74\pm 1.33)\times 10^{-10}$~m/s$^2$ for each particular spacecraft. This signal has become known as the Pioneer anomaly; the nature of this anomaly remains unexplained.

Recently new Pioneer 10 and 11 radio-metric Doppler and flight telemetry data became available.  The newly available Doppler data set is significantly enlarged when compared to the data used in previous investigations and is expected to be the primary source for the investigation of the anomaly.  In addition, the flight telemetry files, original project documentation, and newly developed software tools are now used to reconstruct the engineering history of both spacecraft. With the help of this information, a thermal model of the Pioneer vehicles is being developed to study possible contribution of thermal recoil force acting on the two spacecraft.  The ultimate goal of these physics engineering efforts is to evaluate the effect of on-board systems on the spacecrafts' trajectories.

\end{abstract}


\maketitle

\section{Introduction}

The first spacecraft to leave the inner solar system \citep{JPL1998,JPL2002,JPL2005}, Pioneers 10 and 11 were designed to conduct an exploration of the interplanetary medium beyond the orbit of Mars and perform close-up observations of Jupiter during the 1972-73 Jovian opportunities.

The spacecraft were launched in March 1972 (Pioneer 10) and April 1973 (Pioneer 11) on top of identical three-stage Atlas-Centaur launch vehicles. After passing through the asteroid belt, Pioneer 10 reached Jupiter in December 1973. The trajectory of its sister craft, Pioneer 11, in addition to visiting Jupiter in 1974, also included an encounter with Saturn in 1979 (see \citep{JPL2002,MDR2005} for more details).

After the planetary encounters and successful completion of their primary missions, both Pioneers continued to explore the outer solar system. Due to their excellent health and navigational capabilities, the Pioneers were used to search for trans-Neptunian objects and to establish limits on the presence of low-frequency gravitational radiation \citep{PC202}.

Eventually, Pioneer 10 became the first man-made object to leave the solar system, with its official mission ending in March 1997. Since then, NASA's Deep Space Network (DSN) made occasional contact with the spacecraft. The last successful communication from Pioneer 10 was received by the DSN on 27 April 2002. Pioneer 11 sent its last coherent Doppler data in October 1990; the last scientific observations were returned by Pioneer 11 in September 1995.

The orbits of Pioneers 10 and 11 were reconstructed based primarily on radio-metric (Doppler) tracking data. The reconstruction between heliocentric distances of 20--70 AU yielded a persistent small discrepancy between observed and computed values \citep{JPL2002,JPL2005,MDR2005}. After accounting for known systematic effects \citep{JPL2002}, the unmodeled change in the Doppler residual for Pioneer 10 and 11 is equivalent to an approximately sunward constant acceleration of
\[
a_P = (8.74\pm 1.33)\times 10^{-10}~\mathrm{m/s}^2.
\]

The magnitude of this effect, measured between heliocentric distances of 40--70 AU, remains approximately constant within the 3~dB gain bandwidth of the HGA.  The nature of this anomalous acceleration remains unexplained; this signal has become known as the Pioneer anomaly.

There were numerous attempts in recent years to provide an explanation for the anomalous acceleration of Pioneers 10 and 11. These can be broadly categorized as either invoking conventional mechanisms or utilizing principles of ``new physics''.

Initial efforts to explain the Pioneer anomaly focused on the possibility of on-board systematic forces. While these cannot be conclusively excluded \citep{JPL2002,JPL2005}, the evidence to date does not support these mechanisms: the magnitude of the anomaly exceeds the acceleration that these mechanisms would likely produce, and the temporal evolution of the anomaly differs from that which one would expect, for instance, if the anomaly were due to thermal radiation of a decaying nuclear power source.

Conventional mechanisms external to the spacecraft were also considered. First among these was the possibility that the anomaly may be due to perturbations of the spacecrafts' orbits by as yet unknown objects in the Kuiper belt. Another possibility is that dust in the solar system may exert a drag force, or it may cause a frequency shift, proportional to distance, in the radio signal. These proposals could not produce a model that is consistent with the known properties of the Pioneer anomaly, and may also be in contradiction with the known properties of planetary orbits.

The value of the Pioneer anomaly happens to be approximately $cH_0$, where $c$ is the speed of light and $H_0$ is the Hubble constant at the present epoch. Attempts were made to exploit this numerical coincidence to provide a cosmological explanation for the anomaly, but it has been demonstrated that this approach would produce an effect with the opposite sign \citep{JPL2002,MDR2005}.

As the search for a conventional explanation for the anomaly appeared unsuccessful, this provided a motivation to seek an explanation in ``new physics''. No such attempt to date produced a clearly viable mechanism for the anomaly \cite{MDR2005}.

The inability to explain the anomalous behavior of the Pioneers with conventional physics has resulted in a growing discussion about the origin of the detected signal. The limited size of the previously analyzed data set, also limits our current knowledge of the anomaly. To determine the origin of $a_P$ and especially before any serious discussion of new physics can take place, one must analyze the entire set of radio-metric Doppler data received from the Pioneers.

As of October 2007, an effort to recover this critical information, initiated at JPL in June 2005, has been completed; we now have almost 30 years of Pioneer 10 and 20 years of Pioneer 11 Doppler data, most of which was never used in the investigation of the anomaly. The primary objective of the upcoming analysis is to determine the origin of the Pioneer anomaly.  To achieve this goal, we will investigate the recently recovered radio-metric Doppler and telemetry data focusing on the possibility that the anomaly might have a thermal nature; if so, our analysis will find the physical origin of the effect and will identify its basic properties.

A unique feature of these efforts is the use of telemetry files documenting the thermal and electrical state of the spacecraft. This information was not available previously; however, by May 2006, the telemetry files for the entire durations of both missions were recovered, pre-processed and are ready for the upcoming study. Both of the newly assembled data sets are pivotal to establishing the origin of the detected signal.

In this paper we will report on the status of the recovery of the Pioneers' flight telemetry and its usefulness for the analysis of the Pioneer anomaly.

\section{Using flight telemetry to study the spacecrafts' behavior}

All transmissions of both Pioneer spacecraft, including all engineering telemetry, were archived \citep{MDR2005} in the form of files containing Master Data Records (MDRs). Originally, MDRs were scheduled for limited retention. Fortunately, the Pioneers' mission records avoided this fate: with the exception of a few gaps in the data \citep{MDR2005} the entire mission record has been saved. These recently recovered telemetry readings are important in reconstructing a complete history of the thermal, electrical, and propulsion systems for both spacecraft. This, it is hoped, may in turn lead to a better determination of the spacecrafts' acceleration due to on-board systematic effects.

Telemetry formats can be broadly categorized as science formats versus engineering formats. Telemetry words included both analog and digital values. Digital values were used to represent sensor states, switch states, counters, timers, and logic states. Analog readings, from sensors measuring temperatures, voltages, currents and more, were encoded using 6-bit words. This necessarily limited the sensor resolution and introduced a significant amount of quantization noise. Furthermore, the analog-to-digital conversion was not necessarily linear; prior to launch, analog sensors were calibrated using a fifth-order polynomial. Calibration ranges were also established; outside these ranges, the calibration polynomials are known to yield nonsensical results.

With the help of the information contained in these words, it is possible to reconstruct the history of RTG temperatures and power, radio beam power, electrically generated heat inside the spacecraft, spacecraft temperatures, and propulsion system.

Telemetry words are labeled using identifiers in the form of $C_n$, where $n$ is a number indicating the word position in the fixed format telemetry frames.

\subsection{RTG temperatures and power}

\begin{figure}
\centering \psfig{file=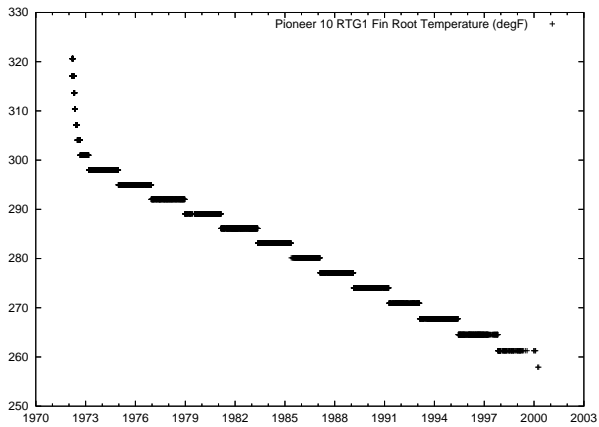, width=0.95\linewidth}
\caption{RTG 1 fin root temperatures (telemetry word $C_{201}$; in $^\circ$F) for Pioneer 10.}
\label{fig:C201}
\end{figure}

The exterior temperatures of the RTGs were measured by one sensor on each of the four RTGs: the so-called ``fin root temperature'' sensor. Telemetry words $C_{201}$ through $C_{204}$ contain the fin root temperature sensor readings for RTGs 1 through 4, respectively. Figure~\ref{fig:C201} depicts the evolution of the RTG 1 fin root temperature for Pioneer 10.

A best fit analysis confirms that the RTG temperature indeed evolves in a manner consistent with the radioactive decay of the nuclear fuel on board. The results for all the other RTGs on both spacecraft are similar, confirming that the RTGs were performing thermally in accordance with design expectations.

RTG electrical power can be estimated using two sensor readings per RTG, measuring RTG current and voltage. Currents for RTGs 1 through 4 appear as telemetry words $C_{127}$, $C_{105}$, $C_{114}$, and $C_{123}$, respectively; voltages are in telemetry words $C_{110}$, $C_{125}$, $C_{131}$, and $C_{113}$. Combined, these words yield the total amount of electrical power available on board:

\[P_E = C_{110}C_{127} + C_{125}C_{105} + C_{131}C_{114} + C_{113}C_{123}.\]

All this electrical power is eventually converted to waste heat by the spacecrafts' instruments, with the exception of power radiated away by transmitters.

\subsection{Electrically generated heat}

\begin{figure}
\centering \psfig{file=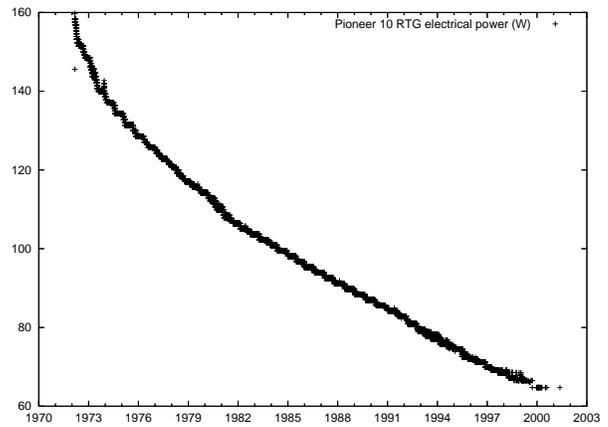, width=0.95\linewidth}
\caption{Changes in total RTG electrical output (in~W) on board Pioneer, as computed using the mission's on-board telemetry.}
\label{fig:elec}
\end{figure}

Whatever remains of electrical energy (Fig.~\ref{fig:elec}) after accounting for the power of the transmitted radio beam is converted to heat on-board. Some of it is converted to heat outside the spacecraft body.

The Pioneer electrical system is designed to maximize the lifetime of the RTG thermocouples by ensuring that the current draw from the RTGs is always optimal. This means that power supplied by the RTGs may be more than that required for spacecraft operations. Excess electrical energy is absorbed by a shunt circuit that includes an externally mounted radiator plate. Especially early in the mission, when plenty of RTG power was still available, this radiator plate was the most significant component external to the spacecraft body that radiated heat. Telemetry word $C_{122}$ tells us the shunt circuit current, from which the amount of power dissipated by the external radiator can be computed using the known ohmic resistance ($\sim$5.25~$\Omega$) of the radiator plate.

Other externally mounted components that consume electrical power are the Plasma Analyzer ($P_\mathrm{PA} = 4.2$~W, telemetry word $C_{108}$ bit 2), the Cosmic Ray Telescope ($P_\mathrm{CRT} = 2.2$~W, telemetry word $C_{108}$, bit 6), and the Asteroid/Meteoroid Detector ($P_\mathrm{AMD} = 2$~W, telemetry word $C_{124}$, bit 5). Though these instruments' exact power consumption is not telemetered, we know their average power consumption from design documentation, and the telemetry bits tell us when these instruments were powered.

Two additional external loads are the battery heater and the propellant line heaters. These represent a load of $P_\mathrm{LH} = P_\mathrm{BH} = 2$~W (nominal) each. The power state of these loads is not telemetered. According to mission logs, the battery heater was commanded off on both spacecraft on 12 May 1993.

Yet a further external load is the set of cables connecting the RTGs to the inverters. The resistance of these cables is known: it is 0.017~$\Omega$ for the inner RTGs (RTG 3 and 4), and 0.021~$\Omega$ for the outer RTGs (RTG 1 and 2). Using the RTG current readings it is possible to accurately determine the amount of power dissipated by these cables in the form of heat:
\[P_\mathrm{cable} = 0.017(C^2_{114}+C^2_{123}) + 0.021(C^2_{127} + C^2_{105}).\]

After accounting for all these external loads, whatever remains of the available electrical power on board is converted to heat inside the spacecraft. So long as the body of the spacecraft is in equilibrium with its surroundings, heat dissipated through its walls has to be equal to the heat generated inside:

\[P_\mathrm{body} = P_E - P_\mathrm{cable} - P_\mathrm{PA} - P_\mathrm{CRT} - P_\mathrm{AMD} - P_\mathrm{LH} - P_\mathrm{BH},\]
with all the terms defined above.

\subsection{Compartment temperatures and thermal radiation}

\begin{figure*}
\includegraphics[width=\linewidth]{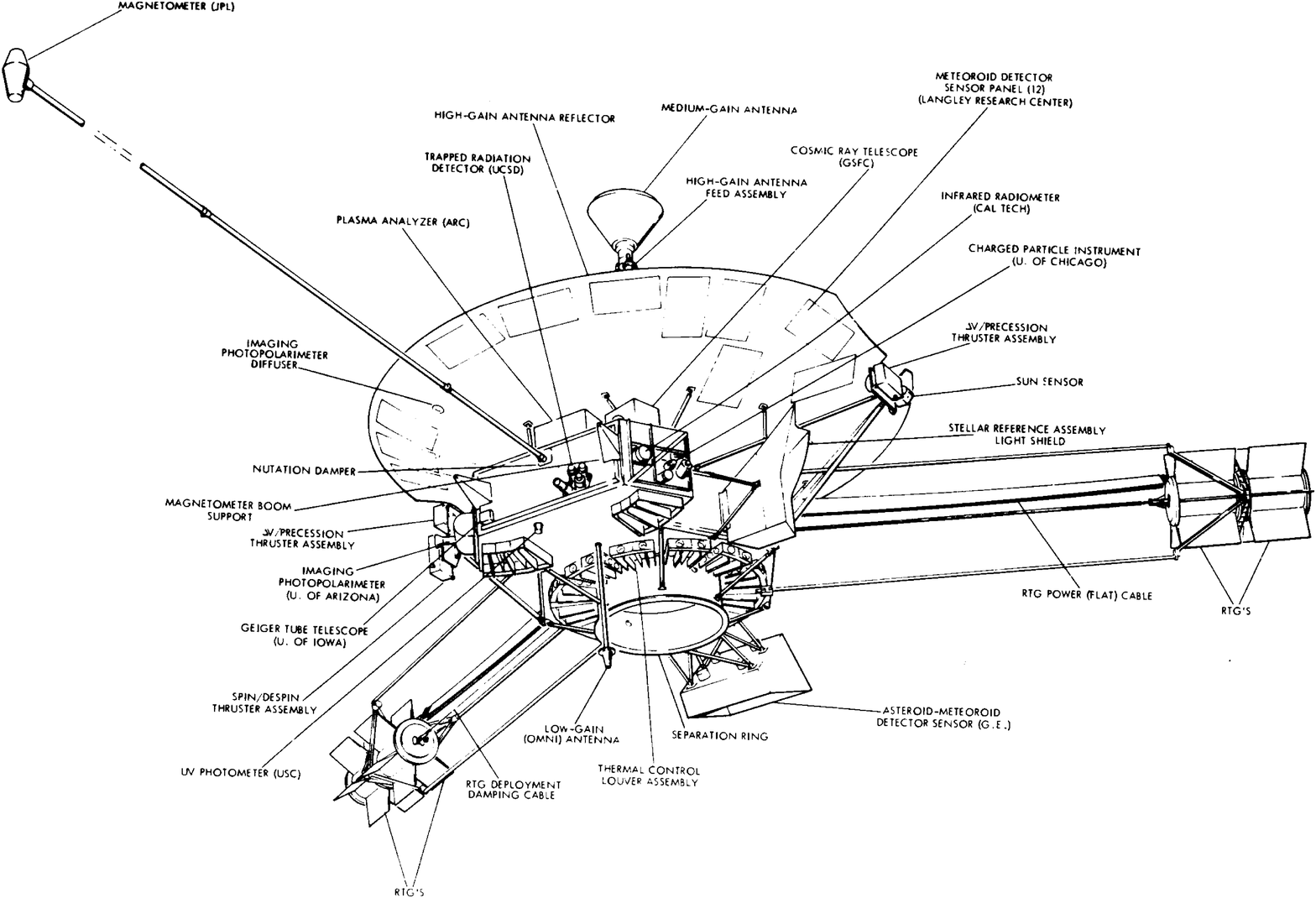}
\caption{A drawing of the Pioneer spacecraft.}
\label{fig:pioneer}
\end{figure*}

As evident from Fig.~\ref{fig:pioneer}, the appearance of the Pioneer spacecraft is dominated by the 2.74~m diameter high gain antenna (HGA). The spacecraft body, located behind the HGA, consists of a larger, regular hexagonal compartment housing the propellant tank and spacecraft electronics; an adjacent, smaller compartment housed science instruments. The spacecraft body is covered by multilayer thermal insulating blankets, except for a louver system located on the side opposite the HGA, which was activated by bimetallic springs to expel excess heat from the spacecraft.

Each spacecraft was powered by four radioisotope thermoelectric generators (RTGs) mounted in pairs at the end of two booms, approximately three meters in length, extended from two sides of the spacecraft body at an angle of 120$^\circ$. A third boom, approximately 6 m long, held a magnetometer.

The total (design) mass of the spacecraft was $\sim$250~kg at launch, of which 27~kg was propellant \citep{PC202}.

For the purposes of attitude control, the spacecraft were designed to spin at the nominal rate of 4.8~rpm. Six small monopropellant (hydrazine) thrusters, mounted in three thruster cluster assemblies, were used for spin correction, attitude control, and trajectory correction maneuvers (see Fig.~\ref{fig:elec}).

\begin{figure}
\includegraphics[width=\linewidth]{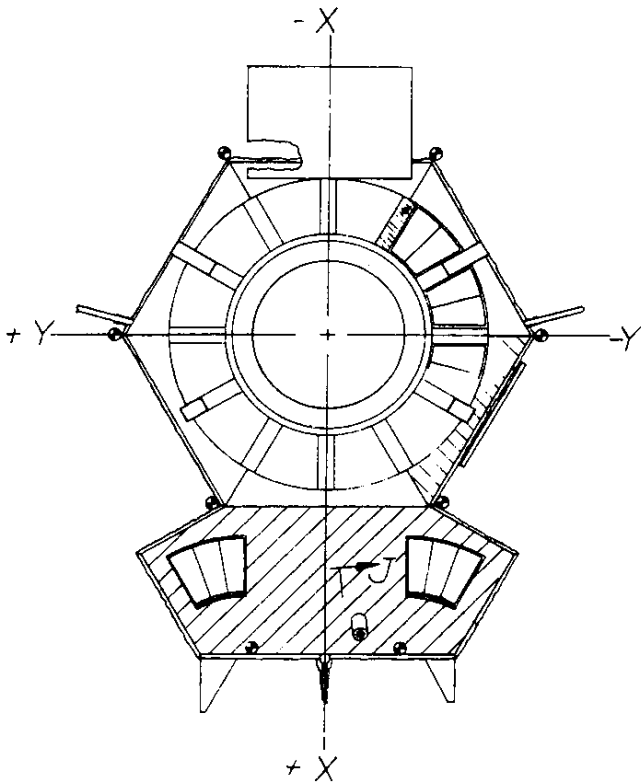}
\caption{Bottom view of the Pioneer 10/11 vehicle, showing the louver system. A set of 12 2-blade louver assemblies cover the main compartment in a circular pattern; an additional two 3-blade assemblies cover the compartment with science instruments.}
\label{fig:louvers}
\end{figure}

The passive thermal control system consisted of a series of spring-activated louvers (see Fig.~\ref{fig:louvers}). The springs were bimetallic, and thermally (radiatively) coupled to the electronics platform beneath the louvers. The louver blades were highly reflective in the infrared. The assembly was designed so that the louvers fully open when temperatures reach 30$^\circ$C, and fully close when temperatures drop below 5$^\circ$C.

The effective emissivity of the thermal blankets used on the Pioneers is $\epsilon_\mathrm{sides} = 0.085$ \citep{MDR2006}. The total exterior area of the spacecraft body is $A_\mathrm{walls} = 4.92$~m$^2$. The front side of the spacecraft body that faces the HGA has an area of $A_\mathrm{front} = 1.53$~m$^2$, and its effective emissivity, accounting for the fact that most thermal radiation this side emits is reflected by the back of the HGA, can be computed as $\epsilon_\mathrm{front} = 0.0013$. The area covered by louver blades is $A_\mathrm{louv} = 0.29$~m$^2$; the effective emissivity of closed louvers is $\epsilon_\mathrm{louv} = 0.04$ \citep{PC202}. The area that remains, consisting of the sides of the spacecraft and the portion of the rear not covered by louvers is $A_\mathrm{sides}= A_\mathrm{walls} - A_\mathrm{front} - A_\mathrm{louv}$.

Using these numbers, we can compute the amount of electrically generated heat radiated through the (closed) louver system as a ratio of total electrical heat generated inside the spacecraft body:

\[P_\mathrm{louver}=\frac{\epsilon_\mathrm{louv}A_\mathrm{louv}P_\mathrm{body}}
{\epsilon_\mathrm{louv}A_\mathrm{louv}+\epsilon_{sides}A_\mathrm{sides}+\epsilon_\mathrm{front}A_\mathrm{front}}\].

This result is a function of the electrical power generated inside the spacecraft body. However, we also have in our possession thermal vacuum chamber test results of the Pioneer louver system. These results characterize louver thermal emissions as a function of the temperature of the electronics platform beneath the louvers, with separate tests performed for the 2-blade and 3-blade louver assemblies. To utilize these results, we turn our attention to telemetry words representing electronics platform temperatures.

\begin{figure}
\includegraphics[width=\linewidth]{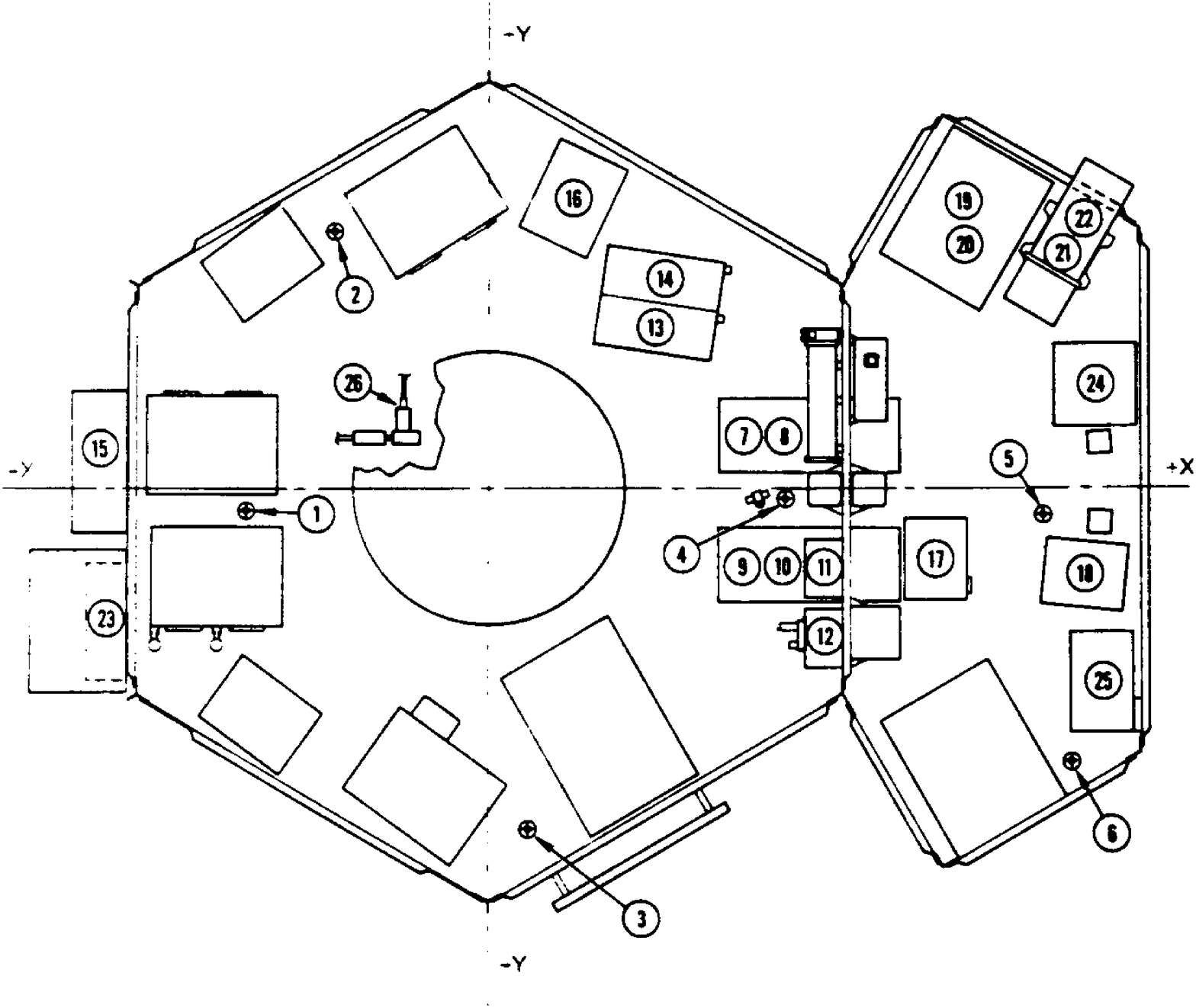}
\caption{Location of thermal sensors in the instrument compartment of Pioneer 10/11 \citep{PC202}. Temperature sensors are mounted at locations 1 to 6.}
\label{fig:tempsens}
\end{figure}

There are 6 platform temperature sensors (Fig.~\ref{fig:tempsens}) inside the spacecraft body: 4 are located inside the main compartment, 2 sensors are in the science instrument compartment. The main compartment has a total of 12 2-blade louver blade assemblies; the science compartment has 2 3-blade assemblies.

The thermal vacuum chamber tests provide values for emitted thermal power per louver assembly as a function of the temperature of the electronics platform behind the louver. This allows us to estimate the amount of thermal power leaving the spacecraft body through the louvers, as a function of platform temperatures \citep{MDR2006}, providing means to estimate the amount of heat radiated by the louver system.

\section{Conclusions}

By 2007, the existence of the Pioneer anomaly is no longer in doubt. A steadily growing part of the community has concluded that the anomaly should be subject to further investigation and interpretation. Our continuing effort to process and analyze Pioneer radio-metric and telemetry data is part of a broader strategy (see discussion at \citep{JPL2005,MDR2005}).

Based on the information provided by the MDRs, we were able to develop a high accuracy thermal, electrical, and dynamical model of the Pioneer spacecraft. This model will be used to further improve our understanding of the anomalous acceleration and especially to study the contribution from the on-board thermal environment to the anomaly.

It is clear that a thermal model for the Pioneer spacecraft would have to account for all heat radiation produced by the spacecraft. One can use telemetry information to accurately estimate the amount of heat produced by the spacecrafts' major components. The next step is to utilize this result along with information on the spacecrafts' design to estimate the amount of heat radiated in various directions.

This entails, on the one hand, an analysis of all available radio-metric data, to characterize the anomalous acceleration beyond the periods that were examined in previous studies. Telemetry, on the other hand, enables us to reconstruct a thermal, electrical, and propulsion system profile of the spacecraft. Soon, we should be able to estimate effects on the motion of the spacecraft due to on-board systematic acceleration sources, expressed as a function of telemetry readings. This provides a new and unique way to refine orbital predictions and may also lead to an unambiguous determination of the origin of the Pioneer anomaly.

\section*{Acknowledgments}

The work of SGT was carried out at the Jet Propulsion Laboratory, California Institute of Technology, under a contract with the National Aeronautics and Space Administration.

\end{document}